\documentclass[reprint,
 amsmath, 
 amssymb, 
prb, 
showkeys, nofootinbib]{revtex4-1}
\usepackage{slashed}
\usepackage{soul}
\usepackage{graphicx}
\usepackage{dcolumn}
\usepackage{bm}
\usepackage{txfonts}
\usepackage[usenames]{color}
\usepackage{appendix}
\usepackage{slashed}
\usepackage[colorlinks=true, citecolor=blue, linkcolor=blue, urlcolor=blue,
pdfpagelabels=true,
hyperindex=true, linktocpage, pdfa=true
]{hyperref}

\usepackage[autostyle]{csquotes}
\usepackage[mathlines]{lineno}
\usepackage[amssymb]{SIunits}
\usepackage{enumitem}

\begin{document}

\preprint{APS/123-QED}

\title{Berry's Phase and Renormalization of Applied Oscillating Electric Fields\\
by Topological Quasi-Particles}

\author{Godwill Mbiti Kanyolo}

\thanks{gmkanyolo@gmail.com; will@inaho.pc.uec.ac.jp}

\altaffiliation{Department of Engineering Science, The University of Electro-Communications, 1-5-1 Chofugaoka, Chofu, Tokyo 182-8585, Japan.\\}


\begin{abstract}
We introduce the concept of Berry's phase in Josephson junctions and consider how this geometric phase arises due to applied oscillating electric fields. The electromagnetic field excites topological quasi-particles from the junction vacuum which affect Cooper-pair tunneling across the Josephson junction barrier. A finite Berry's phase can be detected by its renormalization of the electric field amplitude absorbed by the junction. This has implications for the designing of accurate Josephson junction microwave detectors. 
\end{abstract}

\maketitle


\section{Introduction}
The Berry's phase\cite{Berry1984} is responsible for a wide class of exotic physics ranging from condensed-matter physics and optics to high-energy and particle physics, fluid mechanics to gravity and cosmology.\cite{Berry_app} Here, we consider Berry's phase in Josephson junctions. We show that, a tunneling Cooper pair under the influence of an oscillating electric field can polarize the surface of a tunnel barrier between two superconductors (Josephson junction) leading to topological quasi-particles. These quasi-particles are responsible for an additional phase of topological origin. This phase is responsible for the renormalization of the amplitude of the electric field absorbed by the junction, with implications for microwave detection by Josephson junctions. 

Throughout the paper, we set the Swihart velocity,\cite{Swihart1961} Planck's constant and Boltzmann constant to unity throughout the paper: $\bar{c} = \hbar = k_{\rm B} = 1$. 

\section{Berry's Phase}\label{Intro}

\subsection{Adiabatic Evolution}

When the wavefunction of a quantum system undergoes adiabatic\cite{Adiabatic1928} evolution satisfying the time-dependent Schr{\"o}dinger equation,
\begin{align}
i\frac{\partial}{\partial t}|\psi (t)\rangle = E(t)|\psi(t)\rangle,
\end{align}
the adiabatic wavefunction $|\psi(t) \rangle$ and the Berry's phase\cite{Berry1984} $\gamma(s)$ are given by 
\begin{align}\label{Berry's_wavefunction_eq}
|\psi(s)\rangle = \exp \left [ -i\int_{0}^{s} dt\,E(t) + i\gamma(s)
 \right ]|\psi(\vec{\lambda})\rangle,\\  
\gamma(s) = i\int_{0}^{s} dt\,\langle \psi(\vec{\lambda})|\frac{\partial}{\partial t}|\psi(\vec{\lambda})\rangle,
\end{align}
respectively where the wavefunction $\langle \psi(\vec{\lambda})|\psi(\vec{\lambda}) \rangle = 1$ is normalized. 
The Berry's phase $\gamma$ is observable when it is gauge invariant -- this is the case when the wave function $|\psi[\vec{\lambda}(t)]\rangle$ depends on time via a parameter $\vec{\lambda}(t)$ and $s = T$ is the period of the adiabatic evolution $\vec{\lambda} (t + T) = \vec{\lambda}(t)$, 
\begin{multline}
\gamma (T) = i\int_{0}^{s = T} dt\,\langle \psi(\vec{\lambda})|\frac{\partial}{\partial t}|\psi(\vec{\lambda})\rangle = i\oint d\vec{\lambda}\cdot \langle \psi(\vec{\lambda})|\frac{\partial}{\partial \vec{\lambda}}|\psi(\vec{\lambda})\rangle\\
= \oint d\vec{\lambda}\cdot \vec{\Gamma} = \int d\mathcal{A}\,\vec{n}\cdot\left (\frac{\partial}{\partial \vec{\lambda}}\times\vec{\Gamma} \right ) = \int d\mathcal{A}\,\vec{n}\cdot\mathcal{\vec{B}},
\end{multline}
$\vec{\Gamma} = i\langle \psi(\vec{\lambda})|\frac{\partial}{\partial \vec{\lambda}}|\psi(\vec{\lambda})\rangle$ is the Berry's connection, $\mathcal{\vec{B}} = \frac{\partial}{\partial \vec{\lambda}}\times\vec{\Gamma}$ the Berry's curvature and $\vec{n}$ is the unit vector normal to a surface $\mathcal{A}$ in parameter space $\vec{\lambda}$. 

\subsection{Quantum Phase Dynamics of a Josephson Junction and Aharonov-Bohm Phase}

The quantum phase dynamics for the large Josephson junction\cite{Josephson1962, Large_JJ} is given by,  
\begin{subequations}\label{phase_dynamics_eq}
\begin{align}
eV \equiv \frac{\partial \phi}{\partial t} = 2ed_{\rm eff}\vec{n}\cdot\vec{E},\\
\vec{\nabla}\phi = 2ed_{\rm eff}\vec{n}\times\vec{B},\\
\vec{n}\cdot\vec{J} = 2eE_{\rm J}\sin \phi,
\end{align}
where $e$ is the electric charge, $d_{\rm eff}$ is the effective thickness of the tunnel barrier, $\vec{n}$ is a unit vector normal to the tunnel barrier, $\vec{\nabla} = (\partial/\partial x, \partial/\partial y, \partial/\partial z)$ and $\vec{E}, \vec{B}$ are the electric and magnetic fields respectively which satisfy Maxwell's equations,
\begin{align}\label{Maxwell_J_eq}
\vec{\nabla} \times \vec{B} = \frac{\vec{J}}{\varepsilon_{0}\varepsilon_{\rm r}} + \frac{\partial \vec{E}}{\partial t}. 
\end{align}
\end{subequations}

The quantum phase $\phi$ and thus the vector potential $\vec{A}$ will not vary along the $\vec{n} = (1, \vec{0}_{\vec{n}})$ direction (For simplicity, we assume the junction is oriented in the $x$ direction, thus $\partial \phi/\partial x = 0$). In this case, the quantum phase and electromagnetic fields have generic solutions, 
\begin{subequations}
\begin{align}\label{solutions_eq}
\phi = -2e\int_{0}^{d_{\rm eff}}dx\, \vec{n}\cdot\vec{A} = -2ed_{\rm eff}A_{x},\\
-\frac{\partial A_{x}}{\partial t} = E_{x},\,\frac{\partial A_{x}}{\partial y} = -B_{z},\,\frac{\partial A_{x}}{\partial z} = B_{y}.    
\end{align}
Combining eq. (\ref{phase_dynamics_eq}) with eq. (\ref{Maxwell_J_eq}) leads to the Sine--Gordon equation, 
 \begin{align}
\frac{\partial^2 \phi}{\partial t^2} - \vec{\nabla}_{\vec{n}}^2\phi = -\frac{\vec{n}\cdot \vec{J}}{\varepsilon_{0}\varepsilon_{\rm r}} = -\omega_{\rm p}^2\sin \phi,
\end{align} 
where $\vec{\nabla}_{\vec{n}} = (0, \partial/\partial y, \partial/\partial z)$ and $\omega_{\rm p} = [(2e)^2d_{\rm eff}E_{\rm J}/\varepsilon_{0}\varepsilon_{\rm r}]^{1/2}$ is the Josephson plasma frequency. However, the solution of $\phi$ is not unique since another solution can be obtained by the re-definition, 
\begin{multline}\label{phase_dash_eq}
\phi' = -2e\int_{\mathcal{C},d_{\rm eff}}d\vec{x}\,\cdot\vec{A}\\
= -2e\int_{0}^{d_{\rm eff}}dx\,A_{x} - 2e\oint_{\mathcal{C}} (dy\,A_{y} + dz\,A_{z})\\
= \phi + 2e\int_{\mathcal{A}} dydz\, \left ( \frac{\partial}{\partial z}A_{y} - \frac{\partial}{\partial y}A_{z} \right )\\
= \phi - 2e\int_{\mathcal{A}} dydz\,B_{x},
\end{multline}
\end{subequations}
where $B_{x} = \vec{n}\cdot\vec{\nabla}\times \vec{A}$, $\mathcal{A}$ is the cross-sectional area of the tunnel barrier and $\phi' - \phi$ is the Aharonov-Bohm phase.\cite{Aharonov_Bohm1959} The condition that $\gamma_{\rm AB}$ is undetectable by solely measuring the Josephson current $\vec{n}\cdot\vec{J} = 2eE_{\rm J}\sin \phi$ is $\frac{2e}{2\pi}\int dydz\, B_{x} = k \in \mathbb{Z}$. The tunneling surface $\mathcal{A}$ appears to carry $k$ magnetic charges each carrying a flux quantum, $2\pi/2e$. 

\subsection{The Quasi-Particle Model}

\subsubsection{Renormalization of Oscillating Electric Fields}

Since the 2 dimensional surface of the tunnel barrier acts as a capacitor plate that can store charges, a tunneling Cooper--pair across the barrier and oscillating electromagnetic fields can lead to polarized charges as the capacitor microstates (microscopic degrees of freedom). Consequently, this introduces quantum electro-dynamics in $1 + 2$ dimensions.(See Appendix)

Consider the total Hamiltonian $H = E(t) + U_{\rm int}$ of the Junction including a single quasi-particle of fractional charge 
$2e/k$ and magnetic moment $\vec{\mu}$ interacting adiabatically with an oscillating electric field,
\begin{subequations}\label{Hamiltonian_eq}
\begin{align}
H(t) = \frac{(\vec{P} - 2ek^{-1}\vec{A}^{\rm ac} + \vec{E} \times \vec{\mu})^2}{2M} -2eV - \frac{\partial \gamma}{\partial t}.
\end{align}
Here, $M$, $\vec{P}$ is the mass and momentum of the quasi-particle, $\vec{E} \times \vec{\mu}$ is the Aharonov-Casher potential\cite{Aharonov_Casher1984} acquired by the quasi-particle when it moves around tunneling charges with $\vec{\mu} = -igM^{-1}(2ek^{-1})\,\vec{n}$ a {\it fictitious} imaginary magnetic moment where $g = \beta M$ is the $g-{\rm factor}$ and $\beta$ the inverse temperature, and $-\partial \gamma/\partial t \equiv U_{\rm int}$ is the quasi-particle interaction energy (written in terms of an action $\gamma$) to be determined. 

Under adiabatic evolution due to an oscillating voltage $V(t) = V_{\rm dc} + V_{\rm ac}\cos(\Omega t)$, the total quantum phase $\phi_{\sum}$ of the quasi-particle wavefunction is given by,
\begin{align}
\phi_{\sum}(s) = -\int_{0}^{s}dt\,H(t) = -\int_{0}^{s}dt\,E(t) + \gamma(s). 
\end{align}
\end{subequations}

Since the capacitor couples only to oscillating potentials, the gauge invariant kinetic term in eq. (\ref{Hamiltonian_eq}) will only depend on the time dependent $A_{x}^{\rm ac}$ vector potential, $-\partial A^{\rm ac}_{x}/\partial t = E_{x}^{\rm ac} = d_{\rm eff}^{-1}V_{\rm ac}\cos\Omega t$. Assuming these quasi-particles are topological, the kinetic energy term vanishes\footnote{The energy (density) $E_{\rm kin} \sim \int d^{\,4}x\, T^{00}_{\rm top} = -2\int d^{\,4}x\,\delta\mathcal{L}_{\rm top}/\delta g_{00} = 0$ of topological particles vanishes because their Lagrangian (density) does not depend on the metric tensor $g_{\mu\nu}$\cite{Zee2010}} leading to \begin{align}
\vec{P} = 2ek^{-1}\vec{A}_{\vec{n}}^{\rm ac} + \vec{\mu} \times \vec{E}_{\vec{n}}\\
E(t) = 2eV(t) = 2eV_{\rm dc} + 2eV_{\rm ac}\cos\Omega t = -\frac{\partial \phi}{\partial t}.
\end{align}
Provided the amplitude $2eV_{\rm ac}$ is small compared to $H(t)$, large energy fluctuations are suppressed while the total wave function of the junction $|\psi(\vec{x}_{\vec{n}})\rangle$ undergoes adiabatic evolution leading to,
\begin{subequations}
\begin{multline}\label{renorm_phase_eq}
\gamma(s) = i\int_{0}^{s = 2\pi/\Omega} dt\,\langle \psi[\vec{x}(t)]|\frac{\partial}{\partial t}|\psi[\vec{x}(t)]\rangle\\
= i\int_{d_{\rm eff},\,\mathcal{C}} d\vec{x}\cdot \langle\psi(x)|\vec{\nabla}|\psi(x)\rangle\\
= -\int_{0}^{d_{\rm eff}} dx\,P_{x} - \oint_{\mathcal{C}}d\vec{x}\cdot \vec{P}_{\vec{n}}\\
= -2ek^{-1}d_{\rm eff}A_{x}^{\rm ac} - \oint_{\mathcal{C}}d\vec{x}\cdot \vec{P}_{\vec{n}}. 
\end{multline}
Consequently, we find, 
\begin{multline}
\oint_{\mathcal{C}}d\vec{x}\cdot \vec{P}_{\vec{n}} = \frac{2e}{k}\oint_{\mathcal{C}} (dy\,A_{y}^{\rm ac} + dz\,A_{z}^{\rm ac}) + \frac{2e}{k}i\beta\oint_{\mathcal{C}} (dy\,E_{z} - dz\,E_{y})\\
= \frac{2e}{k}\int dydx\, B_{x}^{\rm ac} +\frac{2e}{k}i\beta\int dydz\,\vec{\nabla}_{\vec{n}}\cdot \vec{E}_{\vec{n}}\\
= \gamma_{\rm AB}^{\rm ac} + \gamma_{\rm AC}.    
\end{multline}
\end{subequations}
Note that because of the term $2ek^{-1}d_{\rm eff}A_{x}^{\rm ac} = k^{-1}\phi_{\rm ac}$ in eq. (\ref{renorm_phase_eq}), the ac voltage is renormalized, 
\begin{subequations}
\begin{align}
\phi_{\sum}(s) = -\int_{0}^{s} dt\,E'(t) - \gamma_{\rm AB}^{\rm ac} - \gamma_{\rm AC},\\
E'(t) = 2eV_{\rm dc} + 2e\Xi V_{\rm ac}\cos(\Omega t), 
\label{renorm_energy_eq}
\end{align}
\end{subequations}
where $\gamma_{\rm AB}^{\rm ac} = -\frac{2e}{k}\int dydx\, B_{x}^{\rm ac} = 2\pi$ is the Aharonov--Bohm phase, $\gamma_{\rm AC} = -\frac{2e}{k}i\beta\int dydz\,\vec{\nabla}_{\vec{n}}\cdot \vec{E}_{\vec{n}}$ the Aharonov--Casher phase and $\Xi = (1 - k^{-1})$.\cite{Aharonov_Bohm1959, Aharonov_Casher1984} 

\subsubsection{Quasi-particle Thermodynamics and Wavefunction Renormalization}

Since it takes $k$ quasi-particles to constitute charge $2e$, this renormalization factor $1 - k^{-1}$ should depend on the statistical average $\langle k \rangle$ at finite temperature. For bosonic quasi-particles, their average number, 
$\langle k \rangle \equiv \langle bb^{\dagger} \rangle = [1 - \exp(-\beta M)]^{-1}$, where $\mathcal{E} \equiv 2M \geq 0$ is the minimum energy required to excite a pair of charged quasi-particles from the vacuum. This leads to a renormalization factor $\Xi^{\rm B} = 1 - \langle k \rangle ^{-1} = \exp(-\beta M)$. On the other hand, fermionic quasi-particles lead to $\langle k \rangle \equiv \langle cc^{\dagger} \rangle = [1 + \exp-\beta M]^{-1}$, where $\Xi^{\rm F} = 1 - \langle k \rangle ^{-1} = -\exp(-\beta M)$. 

Thus, we can rewrite this renormalization factor as $\Xi \equiv \exp(-\beta M) \exp i\beta \omega_{m}$, where $\omega_{m} = \beta^{-1}2m\pi$ or $\omega_{m} = \beta^{-1}(2m +1)\,\pi$ is the bosonic or fermionic Matsubara frequency\cite{Matsubara1955} respectively and $m \in \mathbb{Z}$. Moreover, we are motivated to introduce a Coulomb interaction \footnote{A {\it fictitious} gravitational field $S$ where $\vec{\nabla}_{\vec{n}}S \equiv 2e\beta(\vec{n}\times\vec{E})$} on the $y-z$ plane,
\begin{subequations}
\begin{align}
\vec{\nabla}_{\vec{n}}\cdot \vec{E}_{\vec{n}} \equiv \sum_{l = 1}^{k}\frac{M - i\omega_{m}}{2e}\delta^{\,2}(\vec{r} - \vec{r}_{l}),\\
\gamma_{\rm AC} = k^{-1}2ei\beta\int dydz\,\vec{\nabla}_{\vec{n}}\cdot \vec{E}_{\vec{n}} = i\beta M + \beta\omega_{m},
\end{align}
\end{subequations}
where $\vec{r}_{l} = (y', \,z')$ is the co-ordinate location of the $j^{\rm th}$ quasi-particle on the $y-z$ plane. This leads to a wavefunction renormalization (Lehmann) factor \cite{Lehmann1954, WFR_1951} 
\begin{subequations}
\begin{align}
\psi_{k-1} \rightarrow \psi_{k} = \Xi\,\psi_{k-1},
\end{align}
when $\gamma_{\rm AC} \neq 0$ is finite. Thus, $\Xi = \langle \psi_{k-1}|\psi_{k}\rangle$ is the probability amplitude of exciting the ground state of $\langle k \rangle - 1 = \pm [\exp(\beta M) \pm 1]^{-1}$ quasi-particles with a wavefunction $\psi_{k-1}$ by creating an extra quasi-particle leading to $\langle k \rangle$ quasi-particles with a modified wavefunction $\psi_{k}$. By iteration, we arrive at eq. (\ref{iteration_eq}),
\begin{align}\label{iteration_eq}
\psi_{k} = \Xi\,\psi_{k-1} \cdots \psi_{k\,=\,0} = \Xi^{k}\psi_{k\,=\,0},\\
\langle k \rangle = \sum_{k = 0}^{+\infty}\Xi^{k} \equiv \sum_{k = 0}^{+\infty}\rho_{k}k,
\end{align}
\end{subequations}
and the partition function $\rho_{k} = \Xi^{k}/k = \pm k^{-1}\exp(-\beta kM)$. This yields the quasi-particle free-energy $\mathcal{F} = -\beta^{-1}\ln |\rho_{k}| = \beta^{-1}S + kM$ as expected where $S = \ln k$ is the Boltzmann entropy. 

\subsubsection{Fraction of Absorbed Electric Energy Density}

Since the junction absorbs electromagnetic energy through this process ($\psi_{k-1} \rightarrow \psi_{k}$), we note that due to eq. (\ref{renorm_energy_eq}), the Boltzmann factor $\exp(-\beta 2M) = |\langle \psi_{k}|\psi_{k-1} \rangle|^2$ is the fraction of the electric energy absorbed by the junction. This means that it renormalizes the electromagnetic Lagrangian density of the time-varying periodic potential, 
\begin{subequations}
\begin{align}
\mathcal{L^*}_{\rm M}^{\rm ac} = \frac{\varepsilon_{0}\varepsilon_{\rm r}}{4}\exp(-\beta 2M) E_{x}^{2} + \cdots. 
\end{align}
The reverse ($\psi_{k} \rightarrow \psi_{k - 1}$) relaxation process with $V(t) = V_{\rm dc}$ in eq. (\ref{Hamiltonian_eq}) is accompanied by radiation emission and will lead only to the term $\langle k \rangle^{-1}V_{\rm ac}\cos\Omega t$. Similar arguments yield a factor of $\langle k \rangle^{-1}$ in the Lagrangian.

In particular, introducing the Riemann--Silberstein vector (photon wavefunction)\cite{Maxwell_photon, Riemann_Silberstein} $\vec{\Psi} = k^{-1}\vec{E} + 2i\alpha\vec{B}$ for this process satisfying the vacuum Maxwell's equations, $i\partial \vec{\Psi}/\partial t = \vec{\nabla} \times \vec{\Psi}$, the action to order $\alpha$ is given by
\begin{multline}
S_{\rm ac} = \frac{\varepsilon_{0}\varepsilon_{\rm r}}{4}\int d^{\,4}x\,\vec{\Psi}\cdot\vec{\Psi}\\
= \frac{\varepsilon_{0}\varepsilon_{\rm r}}{4}\int d^{\,4}x\left (k^{-2}\vec{E}\cdot\vec{E} - 4i\alpha\vec{E}\cdot\vec{B} - 4\alpha^2\vec{B}\cdot\vec{B}\right )\\
\simeq \frac{\varepsilon_{0}\varepsilon_{\rm r}}{4k^2}\int d^{\,4}x\,\vec{E}\cdot\vec{E} + \frac{(2e)^2}{4\pi k}i\int  d^{\,4}x\,\vec{E}\cdot\vec{B} - O(\alpha^2),
\end{multline}
\end{subequations}
where $\alpha = (2e)^2/4\pi \varepsilon_{0}\varepsilon_{\rm r}$ is the fine structure constant, $\frac{(2e)^2}{4\pi k}i\int  d^{\,4}x\,\vec{E}\cdot\vec{B} = \frac{(2e)^2}{4\pi k}\int d\tau dydz\,\varepsilon^{\mu\nu\sigma}A_{\mu}\partial_{\nu}A_{\sigma} = S_{\rm CS}$ is a Chern--Simons (topological) term with $d\tau = -idt$ and $k \in \mathbb{Z}$ the level.\cite{CSPontryagin2004,Chern_Simons1982, Zee2010} This suggests that the photon acquires a topological mass $\propto k^{-1}$ when these quasi-particles are present. The significance of $S_{\rm CS}$, $\gamma_{\rm AC}$ and $\gamma_{\rm AB}$ is further discussed in the Appendix. 

\section{Discussion}

\begin{figure}[t!]
\begin{center}
\includegraphics[width=0.6\columnwidth,clip=true]{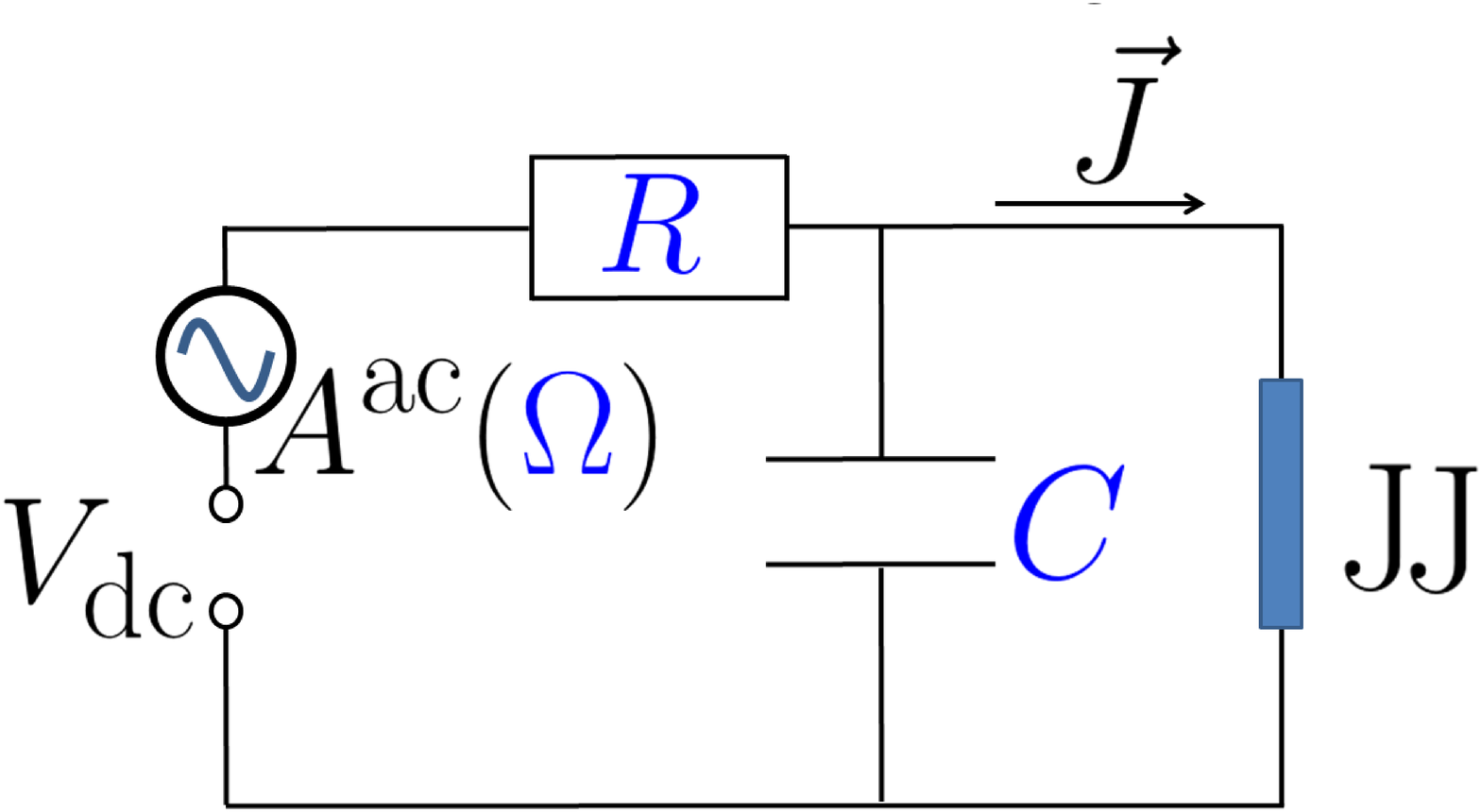}
  \caption{Josephson junction (JJ) equivalent circuit: it yields the simplest response function  $\chi(t) = (1/RC)\exp(-t/RC)$ with $\Xi(\Omega) = [1 + i\Omega RC]^{-1}$.}
\label{circuit_expl}
\end{center}
\end{figure}

The aforementioned renormalization can be discussed within the context of linear response theory. \cite{Kubo1957} Renormalization requires that the applied electric field $E_{x}(t)$ act as an external force, while the renormalized electric field $E'_{x}(t)$ as the causal linear
response of the circuit,\cite{Kanyolo2019_2}
\begin{subequations}
\begin{align}
E'_{x}(t)  = \int_{-\infty}^{t} \chi(t - s) E_{x}(s)ds,
\end{align}
\end{subequations}
where $E'_{x}(\omega) = \Xi(\omega)E_{x}(\omega)$, $E_{x}(\omega) = d_{\rm eff}^{-1}V_{\rm ac}\frac{1}{2}[\delta(\omega - \Omega) + \delta(\omega + \Omega)]$ is the spectrum of the electric field and $\chi(t-s)$ is the response function, making $\Xi(\omega) = \int_0^{+\infty}\chi(t)\exp (-i\omega t) dt$ the susceptibility. Thus, the renormalized electric field spectrum is given by
\begin{subequations}
\label{E_dash_RF_eq}
\begin{align}
E'_{x}(\omega) =  d_{\rm eff}^{-1}V_{\rm ac}\Xi(\omega)[\delta(\omega - \Omega) + \delta(\omega + \Omega)],\\
\Xi(\omega) = |\Xi(\omega)|\exp(i\beta\omega_{m}).
\end{align}
\end{subequations}
For instance, in the equivalent circuit of the Josephson junction (${\rm JJ}$) depicted in Fig. \ref{circuit_expl}, the response function becomes $\chi(t) = (1/RC)\exp(-t/RC)$ with $\Xi(\Omega) = [1 + i\Omega CR]^{-1}$\cite{Grabert2015} and $RC$ the relaxation time of the circuit. Since the Matsubara frequency in this case is given by $\omega_{m} = \beta^{-1}m\pi = \beta^{-1}\arctan(\Omega RC)$, we discover that the oscillation period of the electric field $2\pi \Omega^{-1} \equiv T \gg 2\pi RC$ satisfies the slow (adiabatic)\cite{Adiabatic1928} condition required for the existence of the Berry's phase. 

In conclusion, we have described a model where topological quasi-particles created by an oscillating electric field applied to the tunnel junction renormalizes the amplitude of the electric field via Berry's phase. Since Berry's phase $\gamma(T)$ neither depends on $B_{y}$ nor $B_{z}$, renormalization will be present even in ultra-small junctions ($B_{y} = B_{z} = 0$) exhibiting dynamical Coulomb blockade,\cite{Falci1991, Ingold_Nazarov1992, Grabert2015} as long as the junction responds linearly as discussed. The insights herein are particularly useful in improving the accuracy of ${\rm JJ}$ microwave detectors.\cite{Liou2014, Kanyolo2019} The author would like to thank Prof. Takeo Kato and Prof. Hiroshi Shimada for their constructive critic of the manuscript. Special thanks to Prof. Titus Masese for proofreading the manuscript.

\bibliography{Berry.bib}

\appendix

\section{The Topological Nature of the Quasi-Particles}

\subsection{\label{App: Chern_Simons} Chern-Simons theory}

We can further discuss the significance of $\gamma_{\rm AB}$ and $\gamma_{\rm AC}$ by introducing the Chern--Simons action,\cite{Chern_Simons1982}
\begin{align}\label{Chern_Simons_eq}
S_{\rm CS} = 2e\int dtdydz\,\left (\frac{2e}{4\pi k} \varepsilon^{\mu\nu\sigma}A_{\mu}\partial_{\nu}A_{\sigma} - j^{\mu}A_{\mu} \right ),
\end{align}
where\footnote{Einstein summation convention is used} $\varepsilon^{\mu\nu\sigma}$ is the Levi-Civita symbol in $1 + 2$ dimensions. Since the tunnel barrier acts as a capacitor that can store quasi-particles of charge $2e/k$ on its surface, a tunneling Cooper--pair will lead to a net topological current in the $y-z$ direction. For instance, a Chern-Simons term (a mass for the photon\cite{Chern_Simons1982}) can arise by introducing an effective quantum electrodynamics,\cite{Zee2010}
\begin{multline}\label{S_CS_eq}
\exp i\left [ S_{\rm M} \pm S_{\rm CS} \right ] \sim \\
\int \prod_{n = 1}^{k}D[\bar{\psi}_{n},\psi_{n}, a]\exp i\int d^{\,4}x\,\left [-\frac{\varepsilon_{0}\varepsilon_{\rm r}}{4} f_{\mu\nu}^2\right ]\times\\
\exp i\int dtdydz\,\left [ \bar{\psi}_{n}(i\slashed{\partial} - 2e\slashed{a} - M)\psi_{n} - 2ej^{\mu}A_{\mu}\right ]
\end{multline}
where $S_{\rm M} = -\frac{\varepsilon_{0}\varepsilon_{\rm r}}{4}\int d^{\,4}x\,(\partial_{\mu}A_{\nu}-\partial_{\mu}A_{\nu})^2$ is the Maxwell action, $f_{\mu\nu} = (\partial_{\mu}A_{\nu}-\partial_{\mu}A_{\nu} + \alpha \varepsilon_{\mu\nu\sigma\rho}\partial_{\sigma}a_{\rho})^2$ is the Cabbibo-Ferrari tensor\cite{Fryberger1989} with $\alpha = (2e)^2/4\pi \varepsilon_{0}\varepsilon_{\rm r}$ the fine-structure constant and $\alpha^2 \simeq 0$.

Assuming the current takes the form $j^{\mu} \equiv (\rho, -\beta^{-1}\vec{\nabla}_{\vec{n}}\ln\rho/\rho_{0}) \equiv (\rho, \,\rho\vec{v}\times\vec{n})$ and $\rho$ satisfies the normalization condition $\int dydz\,\rho = 1$, varying $S_{\rm CS}$ with respect to the electromagnetic potential $A_{\mu}$ yields, 
\begin{multline}\label{Chern-Simons_eq}
\delta S_{\rm CS} \rightarrow \frac{2e}{2\pi k}\varepsilon^{\mu\nu\sigma}\partial_{\nu}A_{\sigma} = j^{\nu} \rightarrow \frac{2e}{2\pi k} B_{x} = \rho,\\
\frac{2e}{2\pi k} E_{z} = \beta^{-1}\rho^{-1}\frac{\partial \rho}{\partial y},\,\, \frac{2e}{2\pi k} E_{z} = -\beta^{-1}\rho^{-1}\frac{\partial \rho}{\partial z},  
\end{multline}
where $\frac{2e}{2\pi}\int dydz\, B_{x} = k\int dydz\,\rho = k \in \mathbb{Z}$ is an integer and the superscript -- ${\rm ac}$ -- has been dropped for convenience. 
Thus, the charge density $\rho$ is the source of the magnetic field $B_{x}$ responsible for $\gamma_{\rm AB}$. 

Moreover, since eq. (\ref{Chern-Simons_eq}) guarantees the continuity equation $\partial_{\mu}j^{\mu} = 0$, we can introduce:

1) 1 + 2 dimensional diffusion equation,
\begin{subequations}
\begin{align*}
0 = \partial_{\mu}j^{\mu} \rightarrow \frac{\partial \rho}{\partial t} = \vec{\nabla}_{\vec{n}}\cdot (D\vec{\nabla}_{\vec{n}}\rho),
\end{align*}

2) entropy $S = \ln \rho/\rho_{0}$ and work done $\delta W = \beta^{-1}\delta S$\footnote{The first law of Thermodynamics: $0 = dU = \delta Q - \delta W$ where $\delta$ indicates path dependence of heat $Q$ and work done $W$} by the quasi-particle which need not vanish over a closed path $\mathcal{C}$,   
\begin{align*}
i\gamma_{\rm AC} = 2\pi\beta\oint_{\mathcal{C}} \delta W = 2\pi\oint_{\mathcal{C}} \delta \ln\rho/\rho_{0}
= \frac{2e}{k}\beta\oint_{\mathcal{C}} (dy\,E_{z} - dz\,E_{y}), 
\end{align*}
\end{subequations}
where $D = \beta^{-1}\rho^{-1}$ is the diffusion coefficient satisfying the Einstein–-Smoluchowski relation with $\rho^{-1}$ playing the role of mobility. By inspection, it is clear that eq. (\ref{Chern-Simons_eq}) solves the Euler/Langevin equation\cite{Langevin1997, Euler_eq} below,  
\begin{multline}\label{Langevin_eq}
0 = \frac{d\vec{p}}{dt} = -2\pi\beta^{-1}\vec{\nabla}S
-\frac{2e}{k}(\vec{n}\times\vec{E}) + \frac{2e}{k}(\vec{B}\cdot\vec{n})\vec{n},
\end{multline}
with $-2\pi\beta^{-1} \partial S/\partial x = 2eB_{x}/k$ an entropic force\cite{Verlinde2011} in the $x$ direction. 

\subsection{\label{App: Noise_Geometry} Quasi-Particles as Tori on a 2d surface}

Consider the diffusion operator on a curved 2d surface, 
\begin{align}\label{diffusion_opera}
D\slashed{\nabla}^{2} = \frac{\partial}{\partial t},
\end{align}
where $D$ is the diffusion coefficient, $\slashed{\nabla} = \gamma^{i}\nabla_{i}$, $\partial_{a} = e^{i}_{a}\nabla_{i}$ is a 2d metric compatible covariant derivative, $e^{i}_{a}, e^{a}_{i}$ are (inverse) frame fields that translate between the coordinate frame (metric) $g_{ij} = \delta_{ab}e^{a}_{i}e^{b}_{j}$ and the tetrad frame (metric) $\delta_{ab} = g_{ij}e^{i}_{a}e^{j}_{b}$ with $\delta_{ab}$ the Kronecker delta and $\gamma^{a} = e^{a}_{i}\gamma^{i} = (\sigma_{y}, \sigma_{z})$ are Pauli matrices respectively. The covariant derivative acts on a quasi-particle spinor (fermion) $\psi$ as $\nabla_{i}\psi = (\partial_{i} + i\omega_{i}^{ab}[\gamma_{a},\gamma_{b}]/4)\psi$, where $\omega_{i}^{ab}$ is the spin connection.\cite{Misner2017} 

Defining the average $\langle \hat{O}^2 \rangle = \int d^{\,2}x\sqrt{\det({g_{ij}}})\,(\psi^{*}\hat{O}^2\psi)$ for the differential operator $\hat{O} = D^{1/2}\slashed{\nabla}$, the mean square satisfies,
\begin{subequations}\label{mean_square_eq1}
\begin{multline}
\int_{0}^{\beta} d\tau \langle \hat{O}^2 \rangle 
= \int_{0}^{\beta} d\tau \int d^{\,2}x\sqrt{\det{(g_{ij})}}\,\psi^{*}\partial \psi/\partial t\\
= i\int_{0}^{\beta} d\tau\,\langle \psi|\frac{\partial}{\partial \tau} |\psi\rangle  
= i\oint_{\mathcal{C}} d\vec{x}\cdot \langle \psi|\frac{\partial}{\partial \vec{x}} |\psi\rangle = \gamma_{\rm AB}(\beta),
\end{multline}
where we have used eq. (\ref{diffusion_opera}). This mean-square resembles the Green--Kubo relation $\int_{0}^{+\infty} dt\, \langle \hat{O}(t)\hat{O}(0) \rangle = i\gamma_{\rm AB}$ under the Wick-rotation ($t \rightarrow -i\tau$) where the Aharonov-Bohm phase $\gamma_{\rm AB}(\beta)$ is the transport coefficient\cite{Green1954, Kubo1957} and $\psi(\tau)$ is periodic in $\beta$ (in place of $2\pi/\Omega$) implying that instead the quasi-particles are excited from the vacuum by a heat bath. Nonetheless, for $k$ quasi-particles, $\gamma_{\rm AB} = 2e\oint d\vec{x}\cdot\vec{A} = 2e\int d(Area)\,\vec{n}\cdot\vec{B} = 2\pi k$, which means that $k$ is also a transport coefficient. This is reasonable since $k$ is also the level of the Chern-Simons term in eq. (\ref{Chern_Simons_eq}), related to the conductivity in 2d. Thus, 
\begin{align}
\int_{0}^{\beta} d\tau \int d^{\,2}x\sqrt{\det{(g_{ij})}}\,\psi^{*}\partial \psi/\partial t = 2\pi k
\end{align}
\end{subequations}

Moreover, calculating the mean-square without using eq. (\ref{diffusion_opera}), 
\begin{subequations}\label{mean_square_eq2}
\begin{multline}
\int_{0}^{\beta} d\tau \langle \hat{O}^2 \rangle = \int_{0}^{\beta} d\tau \langle D\slashed{\nabla}^2 \rangle \\
= \frac{1}{2}\int_{0}^{\beta} d\tau\int d^{\,2}x \sqrt{\det(g_{ij})}\, D\psi^{*}\gamma^{i}[\nabla_{i},\nabla_{j}]\gamma^{j}\psi\\
= -\frac{1}{16}\int_{0}^{\beta} d\tau\int d^{\,2}x \sqrt{\det(g_{ij})}\, D\psi^{*}\gamma^{i}\gamma^{j}R_{ijkl}\gamma^{k}\gamma^{l}\psi,   
\end{multline}
where $[\nabla_{i},\nabla_{j}]\psi = -\frac{1}{8}R_{ij}^{\,\,\,\,ab}\gamma_{a}\gamma_{b}\psi$, $R_{ijkl} = K(g_{ik}g_{jl} - g_{il}g_{jk})$ is the Riemann tensor and $K$ the Gaussian curvature of the 2d surface. Substituting $R_{ijkl}$ above, we find
\begin{multline}
-\frac{1}{8}\int_{0}^{\beta} d\tau\int d^{\,2}x\sqrt{\det(g_{ij})}\, KD\psi^{*}\gamma_{i}\gamma_{j}\gamma^{i}\gamma^{j}\psi\\
= \frac{1}{2}\int\,K\sqrt{\det(g_{ij})}\,d^{\,2}x\int_{0}^{\beta} d\tau\,D\psi^{*}\psi\\
= \frac{1}{2}\int\,K\sqrt{\det(g_{ij})}\,d^{\,2}x,
\end{multline}
\end{subequations}
where we have used $\psi^{*}\psi = \rho$ and the Einstein-Smoluchowski relation $D = \beta^{-1}\rho^{-1}$. Comparing eq. (\ref{mean_square_eq1}) with eq. (\ref{mean_square_eq2}), we discover that,
\begin{align}\label{Gauss-Bonnet_eq}
\int\,K\sqrt{\det(g_{ij})}\,d^{\,2}x = 4\pi k = 2\pi\chi,  
\end{align}
which is the Gauss-Bonnet theorem where $\chi = 2 - 2g$ is the Euler characteristic of the 2d surface. Since $\langle k \rangle = [1 + \exp(-\beta M)]^{-1}$, the genus $g$ gives the fermion number $\langle g \rangle = [\exp(\beta M) + 1]^{-1}$. The case for bosons has been discussed in a separate paper.\cite{Kanyolo2020}

\end{document}